\documentclass[showpacs,amsmath,amssymb,aps,10pt,twocolumn]{revtex4-1}
\usepackage[usenames,dvipsnames]{color}
\usepackage{graphicx,textcase}
\usepackage{dcolumn,overpic}
\usepackage{bm,color}
\usepackage[T1]{fontenc}
\usepackage{ae,aecompl}
\usepackage{chngcntr}
\usepackage{color,soul}
\usepackage[para]{threeparttable}
\usepackage{etoolbox}
\usepackage{lipsum}
\AtEndEnvironment{table*}{\vskip10pt}{}{}
\usepackage[bottom]{footmisc}
\usepackage{setspace}
\usepackage{url}
\usepackage{hyperref}
\hypersetup{colorlinks=true,breaklinks,linkcolor=blue,urlcolor=blue,citecolor=blue}

\begin{document}

\title{Magnetic properties of Ruddlesden-Popper phases Sr$_{3-x}$Y$_{x}$(Fe$_{1.25}$Ni$_{0.75}$)O$_{7-\delta}$:  \\
A combined experimental and theoretical investigation}
\author{Samara Keshavarz$^1$, Sofia Kontos$^2$, Dariusz Wardecki$^{3,4,5}$, Yaroslav O. Kvashnin$^1$, Manuel Pereiro$^1$, Swarup K. Panda$^1$, Biplab Sanyal$^1$, Olle Eriksson$^{1,6}$, Jekabs Grins$^3$, Gunnar Svensson$^3$, Klas Gunnarsson$^2$, Peter Svedlindh$^{2,}$}
\email[]{peter.svedlindh@angstrom.uu.se}

\affiliation{$^1$Uppsala University, Department of Physics and Astronomy, Division of Materials Theory, Box 516, SE-751 20 Uppsala, Sweden}
\affiliation{$^2$Uppsala University, Department of Engineering Sciences, Division of Solid State Physics, Box 534, SE-751 21 Uppsala, Sweden}
\affiliation{$^3$Department of Materials and Environmental Chemistry, Stockholm University, S-106 91 Stockholm, Sweden}
\affiliation{$^4$Institute of Experimental Physics, Faculty of Physics, University of Warsaw, Pasteura 5, 02-093 Warsaw, Poland}
\affiliation{$^5$Department of Chemistry and Chemical Engineering, Chalmers University of Technology, Gothenburg, SE-41296, Sweden}
\affiliation{$^6$School of Science and Technology, \"Orebro University, SE-701 82 \"Orebro, Sweden}

\date{\today}

\begin{abstract}
We present a comprehensive study of the magnetic properties of Sr$_{3-x}$Y$_{x}$(Fe$_{1.25}$Ni$_{0.75}$)O$_{7-\delta}$ ($0 \leq x \leq 0.75$). Experimentally, the magnetic properties are investigated using superconducting quantum interference device (SQUID) magnetometry and neutron powder diffraction (NPD). This is complemented by the theoretical study based on density functional theory as well as the Heisenberg exchange parameters. Experimental results show an increase in the N\'eel temperature ($T_N$) with the increase of Y concentrations and O occupancy. The NPD data reveals all samples are antiferromagnetically ordered at low temperatures, which has been confirmed by our theoretical simulations for the selected samples. Our first-principles calculations suggest that the 3D magnetic order is stabilized due to finite inter-layer exchange couplings. The latter give rise to a finite inter-layer spin-spin correlations which disappear above the $T_N$.

\end{abstract}


\maketitle

\section{\label{sec:level1}Introduction}
Materials in the Ruddlesden-Popper (RP) series $A_{n+1}B_{n}O_{3n+1}$, where $A$ is an alkaline-earth or rare-earth element and $B$ is a transition metal element, have attracted considerable interest due to their magnetic properties. Phases based on oxygen deficient strontium ferrate Sr$_{3}$Fe$_{2}$O$_{7-\delta}$ are among the most investigated, where iron will be in a mixed-valence state (Fe$^{3+}$ and Fe$^{4+}$) for $0< \delta <1$, while the oxidation states are Fe$^{3+}$ and Fe$^{2+}$ for $\delta =1$ and $\delta =2$, respectively. The magnetic phase diagram using single crystalline samples of Sr$_{3}$Fe$_{2}$O$_{7-\delta}$ was recently determined for the range $0 \leq \delta \leq1$~\cite{Peets}. The $\delta = 0$ phase, comprised entirely of Fe$^{4+}$ moments, displayed an incommensurate antiferromagnetic (AFM) ordering with a comparably low value for the magnetic ordering temperature ($T_{N} = 115$ K), while the $\delta = 1$ phase, comprised entirely of Fe$^{3+}$ moments, exhibited commensurate AFM ordering with a much higher ordering temperature ($T_{N} \sim 600 $ K). For oxygen deficiencies in between these values, results from magnetometry indicated a more complex magnetic behavior and the magnetic ordering temperature varied in a non-monotonic manner between the values determined for the $\delta = 0$ and the $\delta = 1$ phases. A further complication relates to the layered structure of RP phases and the dimensionality of the magnetic ordering. Neutron powder diffraction (NPD) studies on the RP phases Sr$_{2}$FeO$_{4}$ and Sr$_{3}$Fe$_{2}$O$_{7}$ performed at $4.2$ K showed broad weak magnetic reflections indicative of AFM ordering in two dimensions~\cite{Dann}. These results were contrasted with NPD results obtained at $120$ K for the $\delta = 1$ phase, demonstrating three-dimensional (3D) magnetic ordering. Since inter-layer exchange interaction is expected to be too weak to explain magnetic 3D ordering, it was instead suggested that dipolar coupling between layers was responsible for the observed ordering~\cite{Dann}.

The $\delta = 2$ phase exhibits an $S = 2$ two-legged spin ladder extending along the crystallographic $b$ axis. M\"ossbauer spectra show that AFM order sets in below $T = 296$ K~\cite{Hayashi}, and NPD studies performed at $T = 10$ K indicate a 3D antiferromagnetically ordered state~\cite{Kageyama}. First-principles density functional theory calculations show that the spin-lattice is 2D in terms of exchange interactions and that dipole-dipole interactions are essential for the 3D magnetic ordering of Sr$_{3}$Fe$_{2}$O$_{5}$ at low temperature~\cite{Koo}.

In a previous study, we have investigated the crystal structure and high-temperature properties of the $n=2$ RP Sr$_{3-x}$Y$_{x}$(Fe$_{1.25}$Ni$_{0.75}$)O$_{7-\delta}$ phases ($0 \leq x \leq 0.75$), using X-ray (XRPD) and neutron powder diffraction, thermogravimetry, M\"ossbauer spectroscopy and electrical conductivity measurements~\cite{louise}. Samples as-prepared at $1300$ $^{\circ}$C under N$_{2}$(g) flow as  well as samples subsequently air-annealed at $900$ $^{\circ}$C were studied. RP phases are considered promising candidates for cathode materials in intermediate temperature fuel cells~\cite{Istomin}. The substitution of Sr (Sr$^{2+}$) by Y (Y$^{3+}$) in Sr$_{3}$(Ni,Fe)$_{2}$O$_{7-\delta}$ was motivated by a too large thermal expansion for the fuel cell application and the substitution was expected to decrease the thermal expansion. 
  
In the present study we have investigated the magnetic properties using SQUID magnetometry and NPD of the as-prepared Sr$_{3-x}$Y$_{x}$(Fe$_{1.25}$Ni$_{0.75}$)O$_{7-\delta}$ phases. The experimental work has been complemented with density functional theory (DFT) based first-principles electronic structure calculations, where in addition to magnetic moments also the Heisenberg exchange parameters were evaluated. This paper reports the low temperature AFM structure and the variation of the magnetic transition temperature with Y-content. The smaller ionic radius of Y$^{3+}$ compared to Sr$^{2+}$ implies that the $c$ axis length decreases with increasing Y-content, which is also reflected in an increase of the magnetic ordering temperature. The interest in RP phase materials can emanate from two different standpoints; either from the fact that the material is a high-temperature fuel cell material or from the fact that RP phase materials constitute layered quasi-2D systems with fundamental physics problems to address. In the present combined experimental-theoretical study the focus is on the fundamental understanding of spin ordering in quasi-2D layered systems and how the spin ordering temperature is affected by doping on the A-site. The most intriguing physics problem is the question of the spin ordering dimensionality. Our combined experimental-theoretical study shows that even a weak interlayer exchange interaction is enough to establish 3D spin order at low temperature, but that there will be a dimensional crossover in magnetism at higher temperature.   

\section{\label{sec:level2}Experiment}
\subsection{Materials}
RP phase samples were prepared in the form of Sr$_{3-x}$Y$_{x}$(Fe$_{1.25}$Ni$_{0.75}$)O$_{7-\delta}$ with $0 \leq x \leq 0.75$, through standard solid-state reaction, using SrCO$_{3}$, Y$_{2}$O$_{3}$, Fe$_{2}$O$_{3}$ and NiO as starting materials. Pelletized samples were heated under a N$_{2}$(g) flow at 1300 $^{\circ}$C for 4 $\times$ 17 h, with intermediate grindings. After tempering at 1300 $^{\circ}$C the furnace was cooled down to 700 $^{\circ}$C at a rate of 300 $^{\circ}$C/h and then turned off to cool. Under these conditions the samples contain predominantly Ni$^{2+}$ and Fe$^{3+}$ and the oxygen content is determined by the Y content. The samples have previously been structurally characterized using XRPD, NPD, extended X-ray absorption fine structure (EXAFS), X-ray absorption near edge structure (XANES) and M\"ossbauer spectroscopy~\cite{louise,jekabs}. As discussed in Ref.~\cite{jekabs}, the experimental results indicate that there could be some, $<$20\%, of Fe$^{3+}$ at low $x$-values. Extensive thermogravimetric studies show furthermore that equilibration with the oxygen content in the surrounding atmosphere is very fast at temperatures above 500 $^{\circ}$C, so the oxygen content within each pellet can be expected to be homogeneous.

\begin{figure} [bh]
\includegraphics[width=\columnwidth]{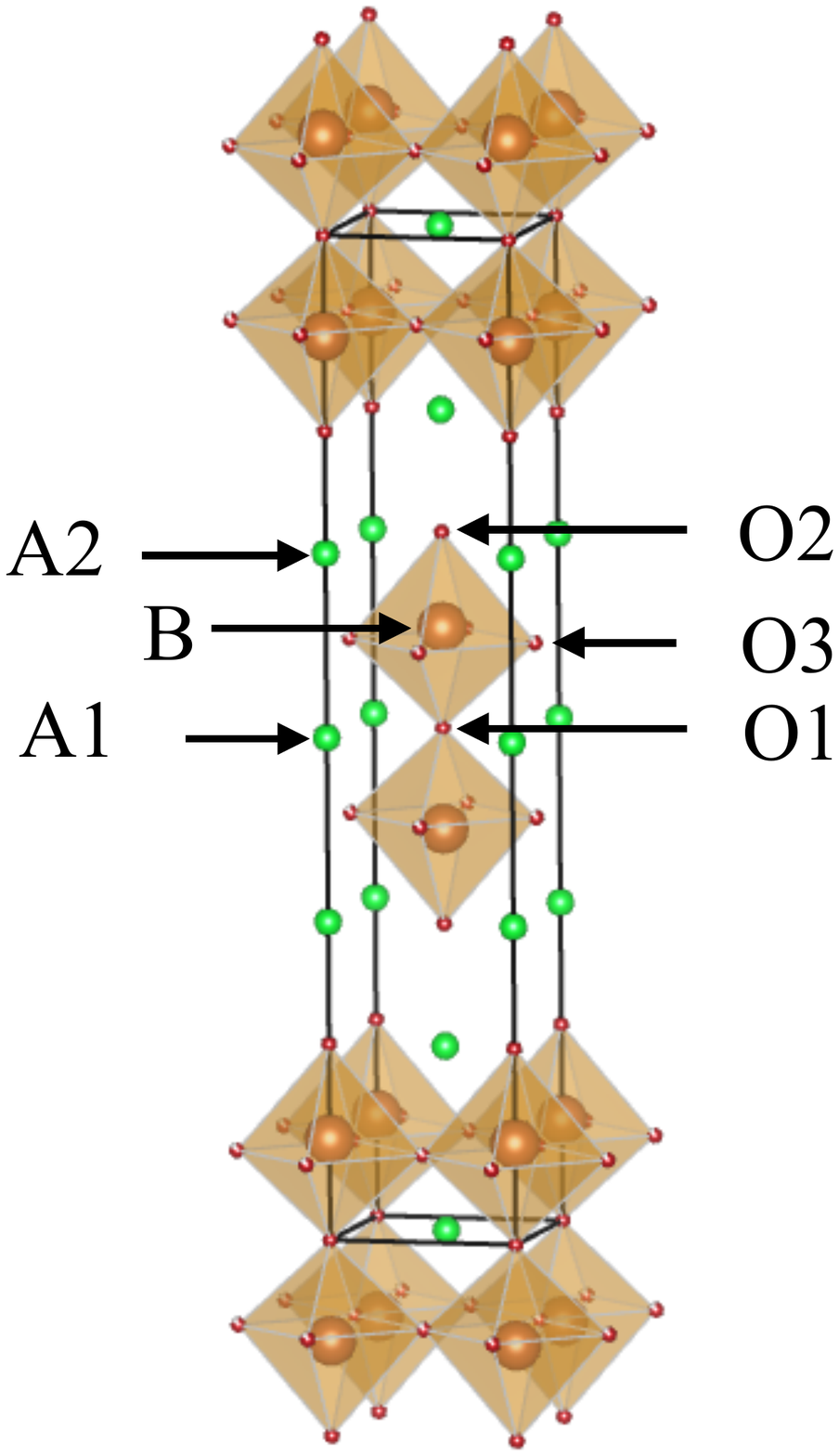}
\caption{Crystal structure of the ideal A$_{3}$B$_2$O$_{7}$ RP structure. There are two different sites for the alkaline-earth or rare-earth elements (A1 and A2) and three different types of oxygen sites (O1, O2 and O3) in the crystal structure. The octahedra surrounding the transition metal elements are shown to visualize the environment of the magnetic elements.}
\label{fig:strucPS}
\end{figure}

The ideal $n=2$ RP structure is shown in Fig.~\ref{fig:strucPS}. There are two different sites for the alkaline-earth or rare-earth elements (A1 and A2) and three different types of oxygen sites (O1, O2 and O3) in the crystal structure. 
In a further study we have investigated the Sr$_{3-x}$Y$_{x}$(Fe$_{1.25}$Ni$_{0.75}$)O$_{7-\delta}$ RP phases by NPD and K-edge Fe and Ni EXAFS/XANES spectroscopy in order to characterize the evolution of the vacancy ordering and oxidation states of Fe and Ni with x, i.e., Y content, and oxygen content in Ref.~\cite{jekabs}. Both as-prepared samples in N2 gas atmosphere and samples subsequently annealed in air at 900$^o$ were characterized. For the as-prepared samples of concern in this study, x = 0.75 has $\delta$ = 1, the O1 atom site vacant, and the Fe$^{3+}$/Ni$^{2+}$ ions have a square pyramidal coordination. With decreasing x the O3 occupancy decreases nearly linearly to 81\% for x =0, while the O1 occupancy increases from 0 for x = 0.4 to 33\% for x = 0. The O2 site is found fully occupied for all compositions. The total oxygen content decreases from 6.0 ($\delta$ = 1) for x = 0.75 to ~ 5.55 ($\delta$ = 1.45) for x = 0. The EXAFS/XANES data show that coordination changes are predominantly for Ni$^{2+}$ ions. The coordination polyhedra for B ions (Fe$^{3+}$ and Ni$^{2+}$ in this work) are therefore square pyramids for x = 0.75, since the O1 site is vacant and the O3 site fully occupied, while with decreasing x in addition to square pyramids also trigonal bipyramid and tetrahedral coordination polyhedra are possible.

\subsection{Methods}
Direct Current (DC) magnetization measurements were performed in a Quantum Design MPMS-XL SQUID magnetometer. Magnetization ($M$) versus temperature ($T$) was studied between $10$ K and $390$ K, following two different protocols; zero-field-cooled (ZFC) and field-cooled (FC). The ZFC magnetization was obtained by cooling the sample to $10$ K in zero field, turning on a weak magnetic field of $H=$ 4 kA/m (50 Oe) and measuring the magnetization as the sample warmed up. The FC magnetization was subsequently obtained by measuring the magnetization, in the same applied field, as the sample cooled down to $10$ K. For the samples \textit{x} = 0.25, 0.40, 0.5, 0.6 and 0.75 the FC magnetization was also measured in a field of $H = 80$ kA/m (1 kOe). Isothermal magnetization measurements were performed at 10 K; the magnetization versus applied field was measured in the field range $\pm$ 4000 kA/m ($\pm$50 kOe).

Time-of-flight (TOF) neutron powder diffraction experiments were performed at both $300$ K and $10$ K using the POWGEN instrument at  SNS in Oak Ridge, US as well as the POLARIS instrument at ISIS, UK. Prior to the diffraction measurements the samples ($1 - 3$ g) were sealed under helium atmosphere in vanadium containers with a diameter of $1$ cm. 

Rietveld analysis of the TOF data performed with the program Topas~\cite{topas} showed that the main nuclear phase for all diffraction patterns can be indexed using a tetragonal cell with {\it I/4mmm} space group symmetry. No nuclear reflections indicative of a super-cell were observed, implying no evidence for a long-range order of the Fe and Ni cations at the B site. However, for some compositions, the patterns showed additional peaks which can be assigned to a small amount ($<$0.3 wt\%) of Ni and NiO. The additional phases were included in the model. 

The TOF diffraction patterns from two banks, covering the range 0.3 \AA\ $< d <$ 14 \AA\ ($\sim$1500 unique reflections) were analysed with a model containing up to 70 free parameters and including the crystal structure taken from Ref.~\cite{jekabs}. In order to improve the fit quality, corrections for anisotropic peak broadening, using 4th order symmetrized spherical harmonics, and absorption were applied. Residual R-factors defined as $R = \sum_{hkl}|I_o - I_c|/\sum_{hkl} I_o$, where $I_o$ and $I_c$ are observed and calculated intensities, respectively, varied with composition in the range 1\% - 5\% for the nuclear phase and 2\% - 5\% for the magnetic phase. The $\chi^2$-values varied from  $\chi^2\approx2$ up to $\chi^2\approx10$ depending on the composition. The atomic thermal vibrations were described by the anisotropic displacement parameters. 

\subsection{Theory}
To better understand the observed magnetic properties of Sr$_{3-x}$Y$_{x}$(Fe$_{1.25}$Ni$_{0.75}$)O$_{7-\delta}$, \textit{ab initio} calculations were performed using the structural parameters of the materials as reported in Ref.~\cite{jekabs}. We performed the calculations for two selected structures characterized by Y concentrations (\textit{x} = 0.5 and 0.75) where the theoretical supercell can sufficiently mimic the experimental systems in terms of the percentage of vacancies and doping. The simulations for other Y concentrations require a larger supercell which would be computationally too demanding. 

The electronic structure as well as the magnetic properties were studied in the framework of DFT~\cite{dft}, using the scalar-relativistic full-potential linear muffin-tin orbital (FP-LMTO) code RSPt~\cite{rspt-web,rspt-book}. Due to the full-potential character, this method does not imply any limitations on the geometry of the systems making it especially suitable to systems with impurities, vacancies as well as non closed-packed structures. The DFT simulations were performed using the local density approximation (LDA) as the exchange-correlations functional~\cite{lda}. For the integration over the Brillouin zone a \textit{k}-point mesh of 11$\times$11$\times$3 has been used. Since plain DFT provides a rather poor description of the localized 3\textit{d} orbitals, the Hubbard $U$ correction has been applied to the set of Fe and Ni 3\textit{d} orbitals. The input parameters in LDA+$U$ calculations are the Coulomb parameter $U$ and the Hund's exchange $J$. We have taken the commonly used values for the $U$ and $J$ from the literature~\cite{fe-u,ni-u} with $U$ = 3 eV for Fe and $U$ = 6 eV for Ni and the value of $J$ was set equal to 0.8 eV for both Fe and Ni. The commonly used double-counting (DC) correction scheme, based on the fully localized limit (FLL) formulation~\cite{dc} was adopted. This DC correction scheme is specially suitable for the systems in which the electrons are rather close to the atomic limit, e.g., insulators like transition metal oxides.

Next, the converged electronic structure of the system has been used to extract the exchange parameters. This is done through mapping the magnetic excitations onto the Heisenberg Hamiltonian
\begin{equation}
\hat{H}=-\frac{1}{2}\sum_{i \neq j }J_{ij}\vec{S}_i\vec{S}_j,
\end{equation}
where $J_{ij}$ is the exchange parameter between two spins, located at sites \textit{i} and \textit{j}, and $\vec{S}_i$ is a vector along the magnetization direction of the spin at the corresponding site. The exchange parameters were computed using a formalism initially suggested in Ref.~\cite{jij-1} and adapted for the current bases set in Ref.~\cite{Jijs-in-rspt}. Note that expressions that relate electronic structure theory to Heisenberg exchange parameters have been also suggested in Refs.~\cite{kore1} and~\cite{kore2}. According to this method, one can extract pair-wise exchange interactions based on the energy changes due to infinitesimal rotations of the spins. Among the most important physical approximation related to the LDA+U Hamiltonian is the shape of the correlated orbitals. In this work, we used MT-heads projection which keeps the radial part of the LMT orbitals inside the muffin-tin sphere and ignores the interstitial part. More specific technical details about the implementation of this method in RSPt as well as the evaluation of the exchange parameters with respect to the choice of the basis set for the localized orbitals are given in Ref.~\cite{Jijs-in-rspt}. The spin-orbit coupling has not been taken into account in the calculations of the exchange parameters. This term introduces, e.g., anisotropic exchange interactions. The bilinear term $J_{ij}$, however, is the leading term determining the magnetic order. Finally, the extracted exchange parameters from the simulations were used for the estimation of the ordering temperatures and the spin-spin correlation function using the classical Monte Carlo method as implemented in the UppASD code~\cite{uppasd-book,uppasd}.

\section{\label{sec:level4}Results and Discussion}

\begin{figure}
\centering
\includegraphics[width=\columnwidth]{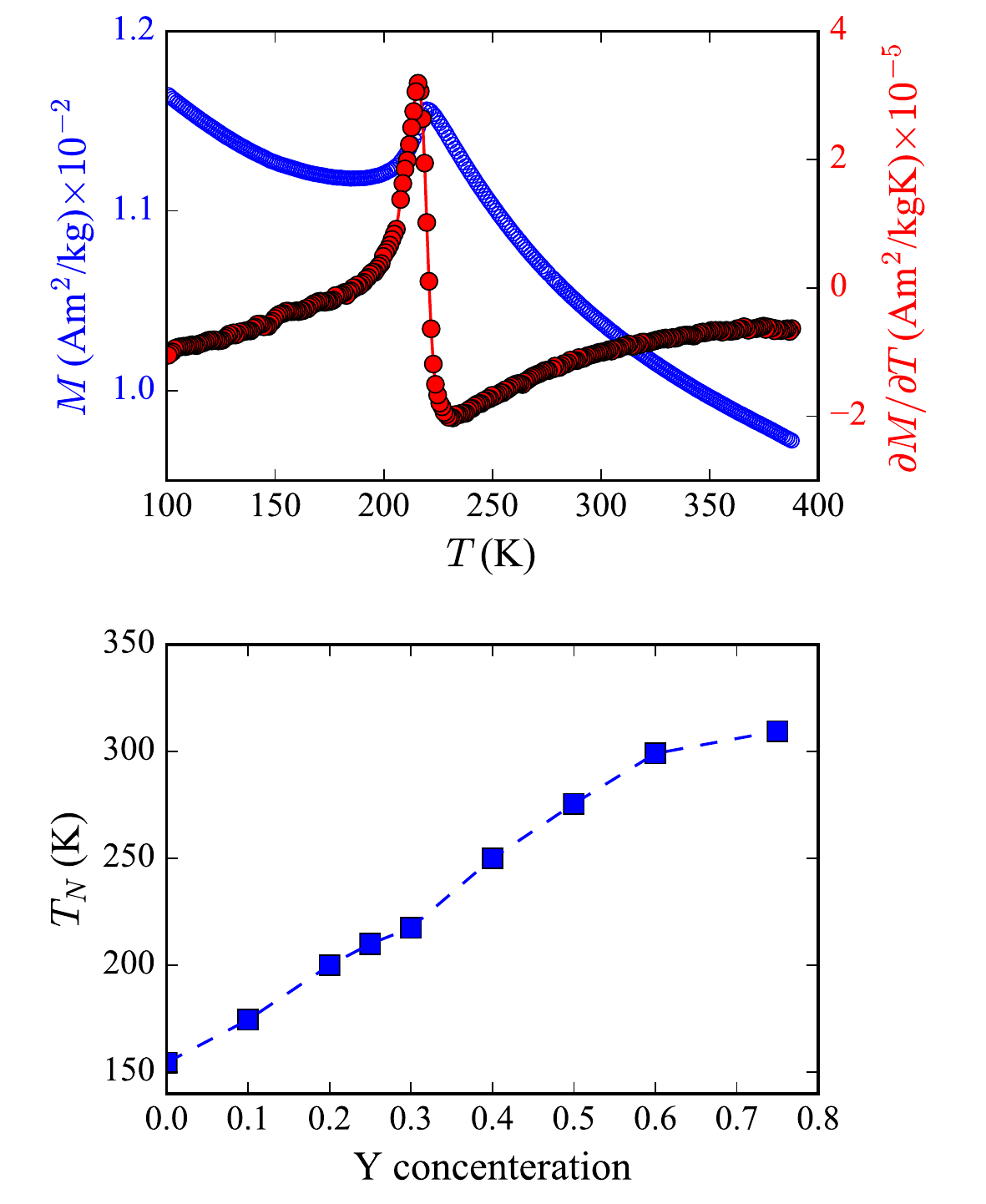}
\caption{Top) Field cooled magnetization $M_{FC}$ (left-hand scale) and $\partial M_{FC}/\partial T$ (right-hand scale) versus temperature $T$ for the $x = 0.3$ sample. The applied magnetic field was $H=$ 80 kA/m. Bottom) $T_N$ versus Y concentration.}
\label{fig:Neel1}
\end{figure}

The results from temperature dependent magnetization measurements show that all samples order antiferromagnetically below a Y concentration dependent ordering temperature ($T_{N}$).  Figure~\ref{fig:Neel1} (a) shows the FC magnetization ($M_{FC}$; left-hand scale)) as well as the temperature derivative of $M_{FC}$ ($\partial M_{FC}/\partial T$; right-hand scale) versus temperature for the $x = 0.3$ sample; $T_{N}$ is defined as the temperature wherein $\partial M_{FC}/\partial T$ exhibits a maximum. It can be seen in this figure that $M_{FC}$ at temperatures below $T_{N}$ increases with
decreasing temperature suggesting a paramagnetic contribution to the measured magnetization, an observation which is more prominent for smaller Y concentrations. Such a paramagnetic contribution has also been identified in results from iron-57 M\"ossbauer spectroscopy studies performed on Sr$_{3-x}$Y$_{x}$(Fe$_{1.25}$Ni$_{0.75}$)O$_{7-\delta}$ samples~\cite{louise}. The M\"ossbauer identified paramagnetic contribution was assigned to paramagnetic Fe$^{3+}$, which due to O3 vacancies can arise from isolated Fe$^{3+}$ ions. In the case of the magnetization results, both isolated Fe$^{3+}$ and Ni$^{2+}$ ions will contribute to the paramagnetic signal. Figure~\ref{fig:Neel1} (bottom panel) shows how $T_{N}$ varies with Y concentration, increasing almost linearly with increasing Y concentration from $x = 0$ and reaching a plateau when approaching $x = 0.75$. The variation of $T_{N}$ with $x$ will be further discussed below in connection with the occupancy of the O1 and O3 sites. 

\begin{figure}[t]
\centering
\includegraphics[width=\columnwidth]{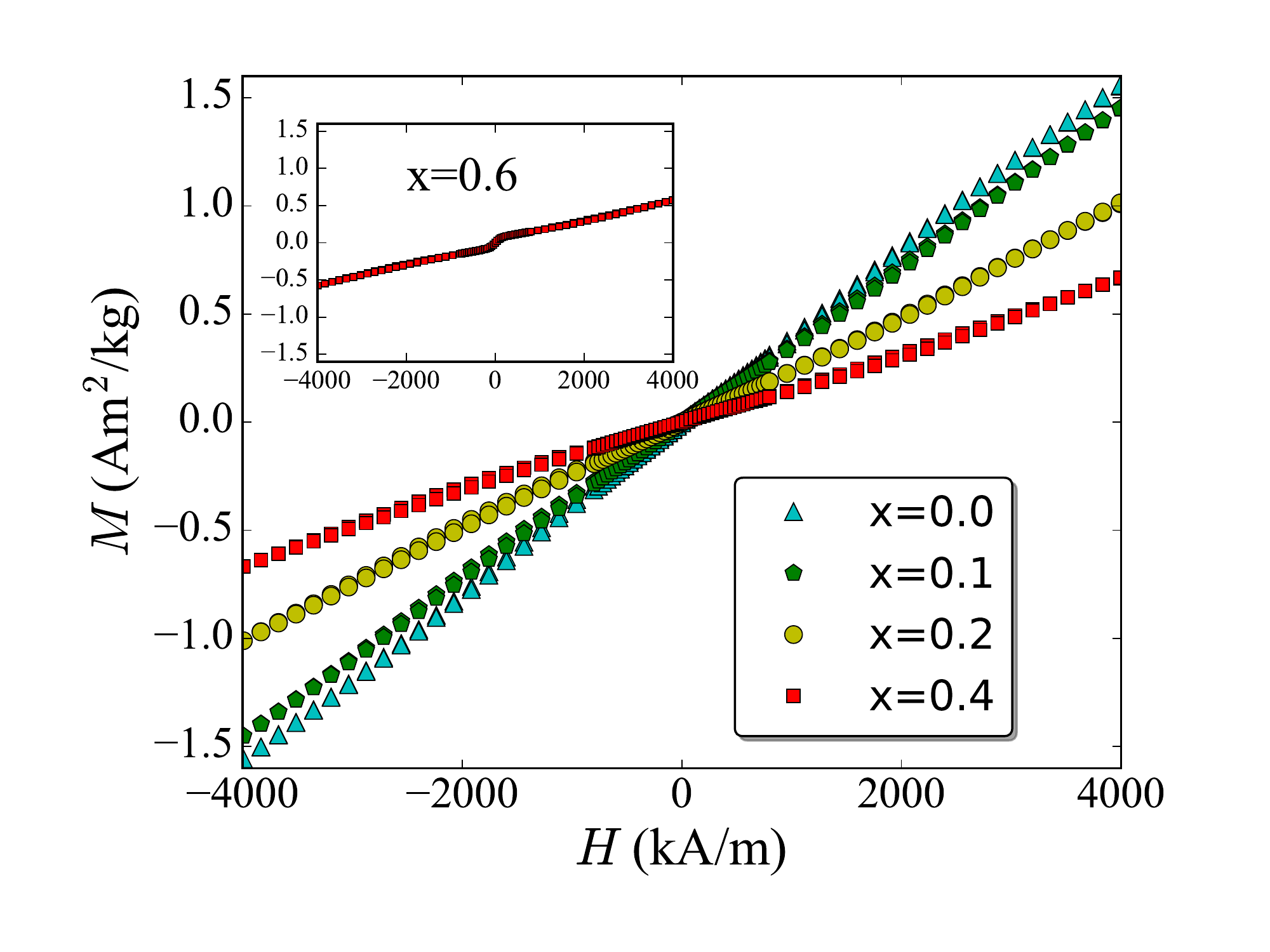}
\caption{Magnetization $M$ versus magnetic field $H$ for samples with different Y concentrations $x$. The insert shows the magnetization curve for the sample with $x = 0.6$, showing a weak anomaly at small fields corresponding to a small amount ($\approx$ 0.25 wt\%) of Ni impurity phase.}
\label{fig:Neel3}
\centering
\includegraphics[width=\columnwidth]{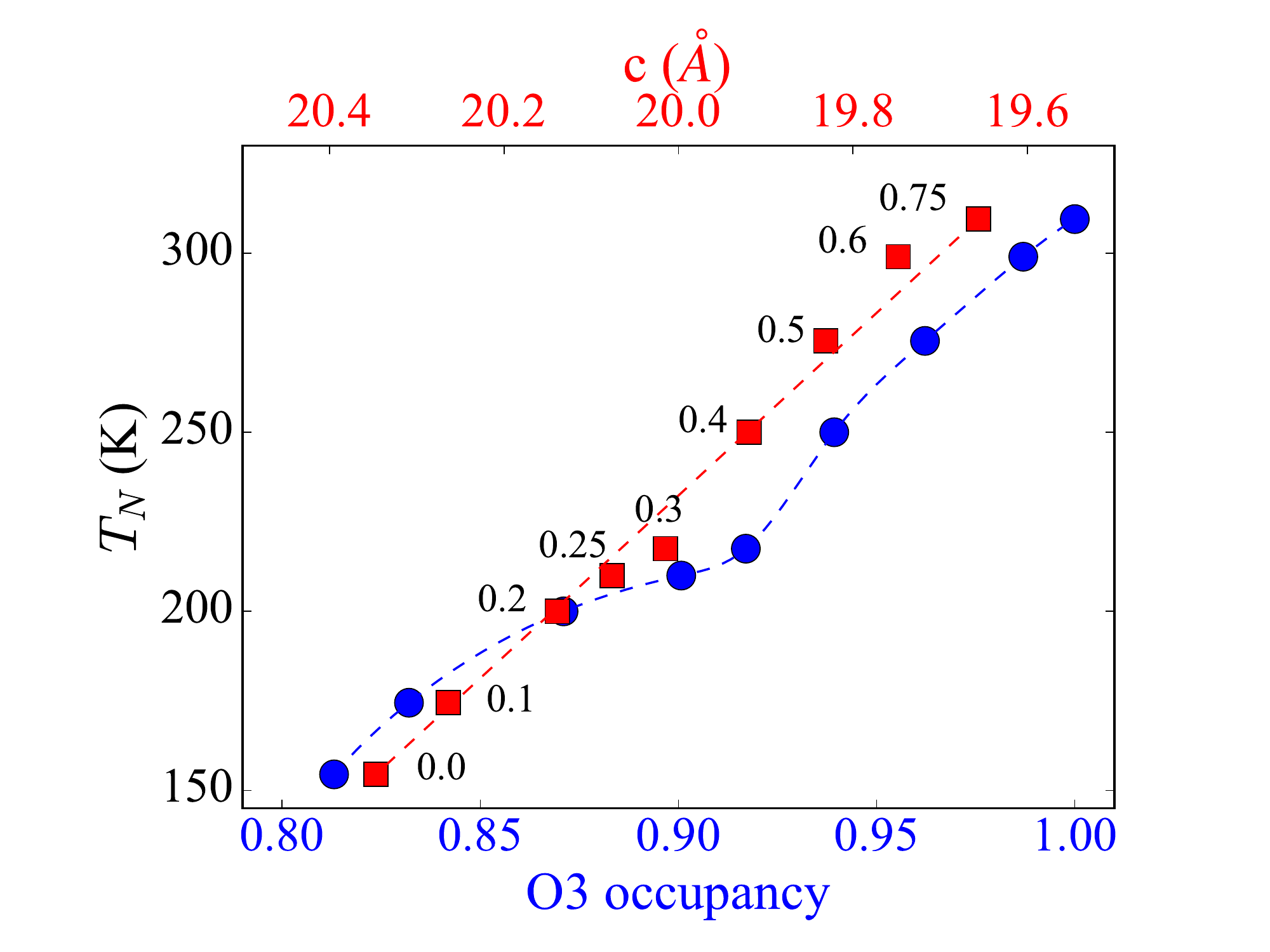}
\caption{$T_N$ versus occupancy of the O3-site (lower scale; blue circles) and $c$-axis length (upper scale; red squares). The number in the plot correspond to the $x$ values in that point.}
\label{fig:Neel2}
\end{figure}

Figure~\ref{fig:Neel3} shows the magnetization versus magnetic field results measured at $T = 10$ K for samples with different Y concentrations. No magnetic hysteresis is apparent, which is consistent with a low temperature antiferromagnetic state for all samples. The magnetic response increases with decreasing $x$, which could be explained by a larger fraction of isolated Fe$^{3+}$ and Ni$^{2+}$ ions for smaller $x$-values, yielding in addition to the antiferromagnetic response a paramagnetic contribution to the measured magnetization. It may also be noted that $\partial M/\partial H$ increases slightly with increasing field. This could indicate that a spin-flop transition will occur at higher fields, as has been observed for other layered systems with $d^5$ ions~\cite{spinflop}. However, the indications of a spin-flop transition in our results are weak and it may be that the spin-flop transition will be smeared out because of a random distribution of Fe$^{3+}$ and Ni$^{2+}$ ions on the B sites. The insert in Fig.~\ref{fig:Neel3} shows the magnetization curve for the $x=0.6$ sample (similar results were obtained for the $x=0.75$ sample). The weak signature at small fields has been identified to originate from a small amount ($\approx$ 0.25 wt\%) of Ni impurity phase. The signature exists up to the highest temperature studied in this work ($T = $ 390 K) and the amplitude follows a temperature dependence that can be expected for the saturation magnetization of Ni. 

\begin{figure}[tp]
\centering
\includegraphics[width=\columnwidth]{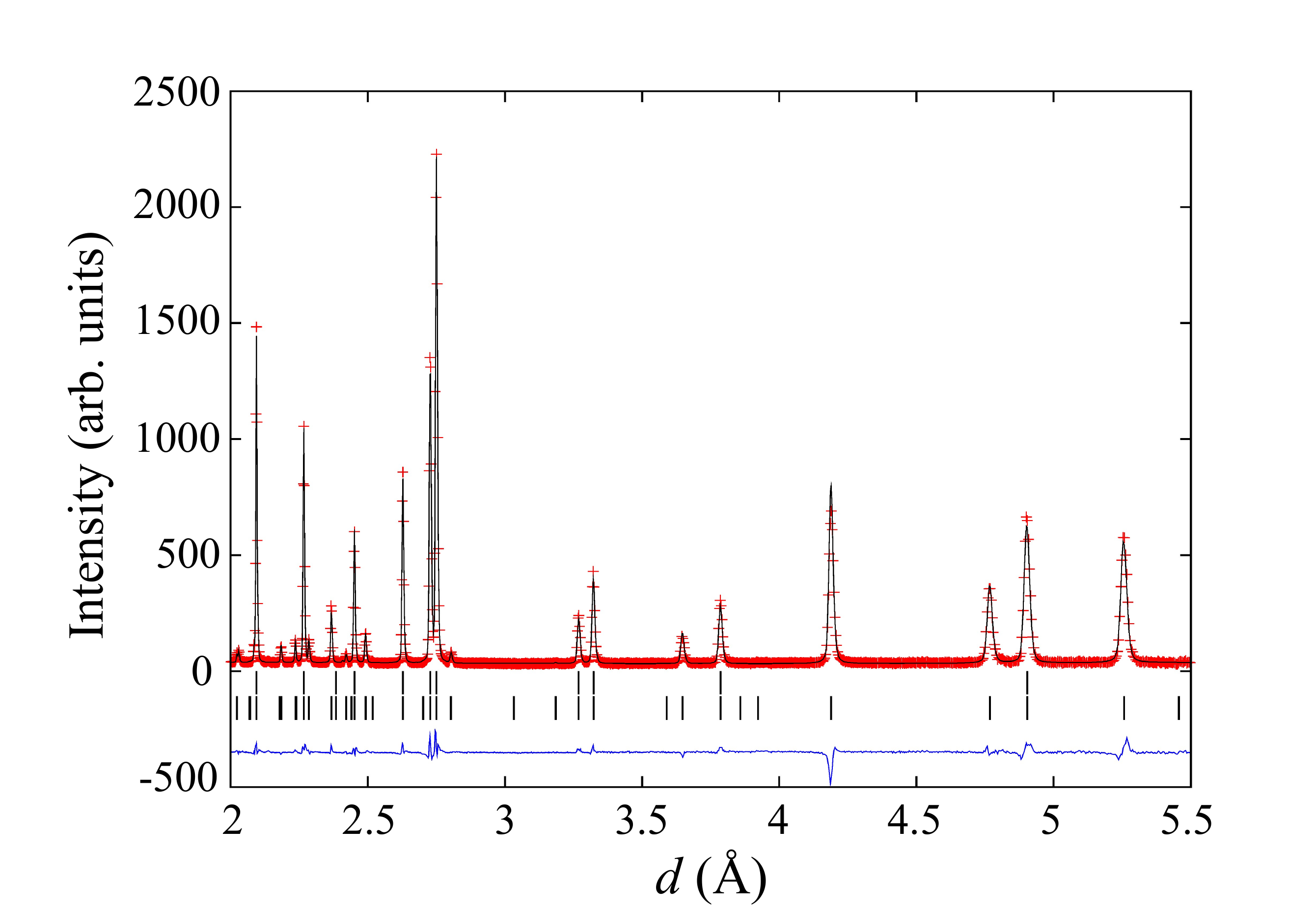}
\caption{Neutron powder diffraction pattern at $T = 10$ K for the $x = 0.75$ sample. Solid lines and symbols indicate measured and calculated patterns, respectively. The difference between observed and calculated patterns is shown at the bottom. The bars indicate the positions of the chemical (top) and magnetic (bottom) Bragg reflections.}
\label{fig:magneticpattern}
\end{figure}

Figure~\ref{fig:Neel2} shows how $T_N$ varies with occupancy of the O3-site (lower scale); the data for the O3-site occupancy is taken from Refs.~\cite{louise, jekabs}. $T_N$ is largest for $x = 0.75$ for which the O3-site is fully occupied and then gradually decreases as the occupancy decreases. The rate of $T_N$-decrease increases down to approximately an O3 occupancy of 0.9, while at even lower O3 occupancy the decrease of $T_N$ proceeds more slowly. This change of trend occurs when the O1-site begins to be occupied; the occupancy of the O1-site raises above zero below $x = 0.4$  and increases with decreasing $x$ reaching $\approx 0.33$ for $x=0$~\cite{louise, jekabs}. Figure~\ref{fig:Neel2} also shows $T_N$ versus the c-axis length; the data for the c-axis length is taken from Refs.~\cite{louise}. The decrease of the c-axis length with increasing Y concentration is expected since the ionic radius for Y$^{3+}$, $r$(Y$^{3+}$)$=1.079$ {\AA}, is smaller than that of Sr$^{2+}$, $r$(Sr$^{2+}$)$=1.31$ {\AA}. $T_N$ is largest for $x = 0.75$ for which the $c$-axis takes its smallest value and then decreases almost linearly with increasing $c$-axis length. The strong correlation between the $T_N$ and the $c$-axis length suggests a corresponding correlation between the interlayer exchange interaction and the $c$-axis length.

Figure~\ref{fig:magneticpattern} shows the measured and calculated NPD patterns at $T = 10$ K for the $x = 0.75$ sample, with chemical (top) and magnetic (bottom) magnetic diffraction peaks indicated by the bars. All samples exhibit similar NPD patterns at $T = 10$ K. The NPD data revealed that all samples are magnetically ordered at 10 K. Moreover, the sample with the largest Y concentration ($x$ = 0.75) also shows magnetic peaks at 300 K. The symmetric appearance of the magnetic Bragg profiles strongly favors an interpretation in terms of 3D magnetic ordering~\cite{bragg-sym}.

\begin{figure}[h]
\centering
\includegraphics[width=\columnwidth]{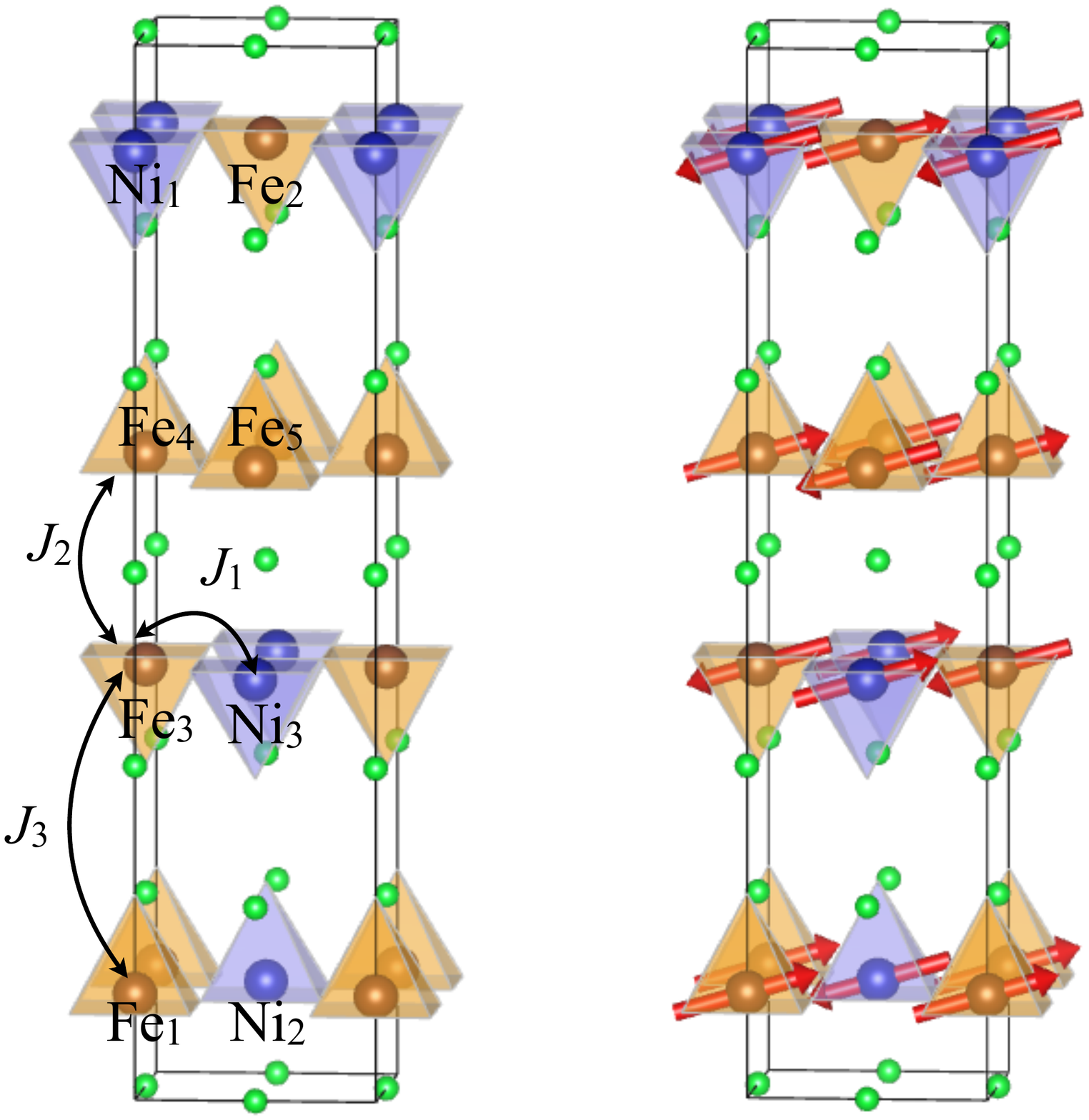}
\caption{Left: Positions of the different 3$d$ atoms in the RP structure for Sr$_{2.25}$Y$_{0.75}$(Fe$_{1.25}$Ni$_{0.75}$)O$_{7-\delta}$ are indicated as well as the intra-layer ($J_1$) and inter-layer ($J_2$ and $J_3$) exchange interactions. Right: Magnetic structure obtained by Rietveld refinement of neutron powder diffraction data.}
\label{fig:Y075}
\end{figure}

In the model of the magnetic structure that gives the best fit, the magnetic moments have two components: in-plane ($\mu_{xy}$) and out-of-plane ($\mu_z$), and the moments of the nearest neighbors are anti-parallel forming a $G$-type antiferromagnetic structure within the perovskite double layer, as shown in right panel of Figure~\ref{fig:Y075}. In other words the magnetic structure is commensurate with the nuclear structure and can, with reference to the nuclear cell, be described by the propagation vector $\vec{\kappa} = (1/2, 1/2, 0)$. The magnetic moments for different $x$ values, measured at 10 K are listed in Table~\ref{tab:mag-exp}. The largest magnetic moment of 3.2 $\mu_B$ is for $x = 0.75$ and it decreases with decreasing $x$.

\begin{table}[t]
    \centering
    \caption{In-plane ($\mu_{xy}$) and out-of-plane ($\mu_z$) magnetic moment components, and total magnetic moment ($\mu_{tot}$) at 10 K extracted from NPD data.}
    \begin{ruledtabular}
    \begin{tabular}{lccc}
            & \multicolumn{3}{c}{magnetic moment $(\mu_B)$}\\
            \cline{2-4}
        $x$ & $\mu_{xy}$ & $\mu_z$ & $\mu_{tot}$ \\
        \hline
         0.0 & 2.19(4) & 1.0(1) & 2.42(4)\\
         0.1 & 2.48(3) & 1.2(1) & 2.74(4)\\
         0.25& 2.52(4) & 0.8(1) & 2.64(5)\\
         0.4 & 2.77(3) & 0.8(1) & 2.89(3)\\
         0.5 & 2.96(3) & 0.8(1) & 3.06(3)\\
         0.6 & 2.96(4) & 1.0(1) & 3.12(4)\\
         0.75& 2.97(2) & 1.2(1) & 3.19(2)\\
    \end{tabular}
    \end{ruledtabular}
    \label{tab:mag-exp}
\end{table}

\begin{figure*}[t]   
\includegraphics[width=\columnwidth]{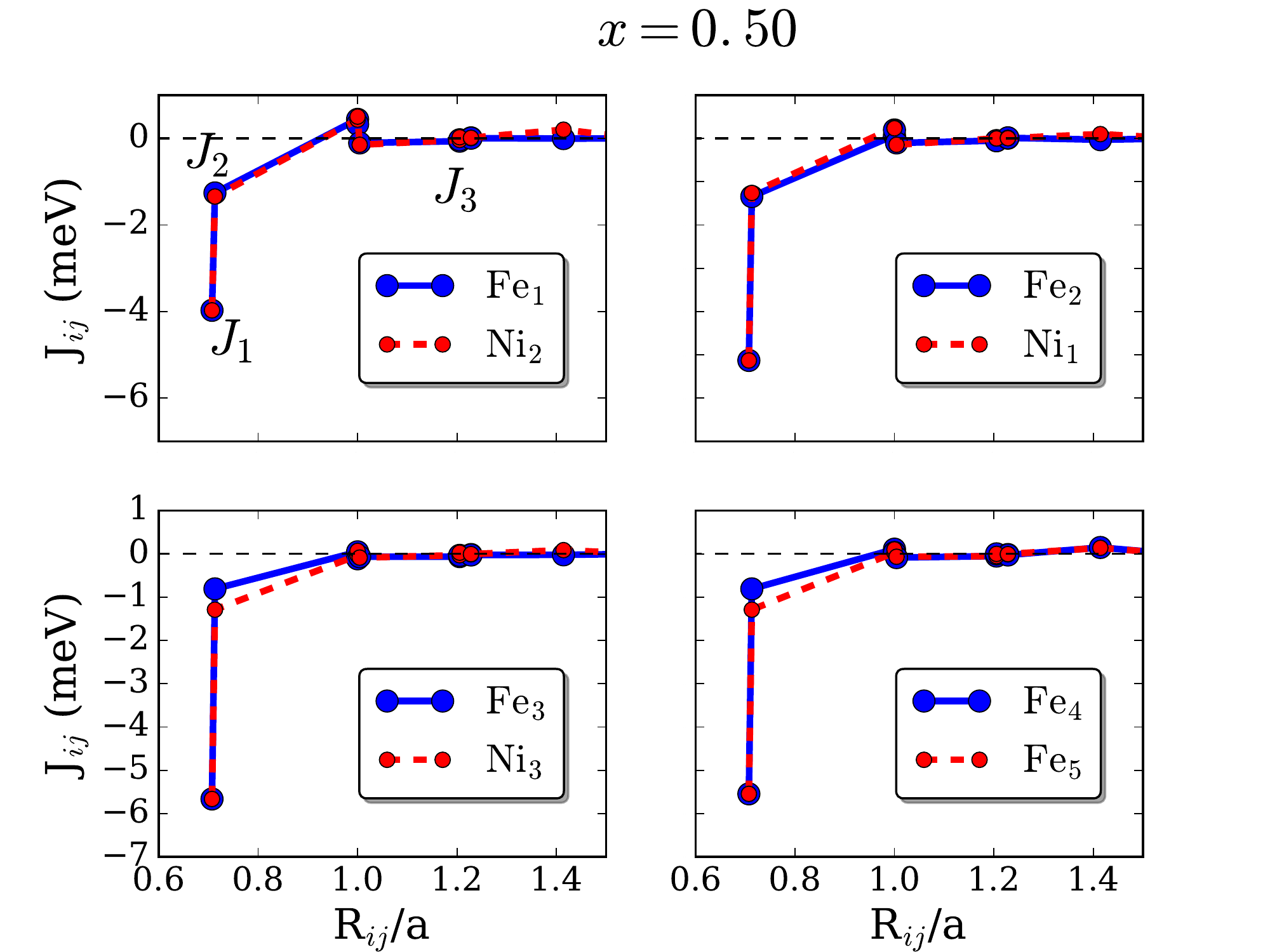} 
\includegraphics[width=\columnwidth]{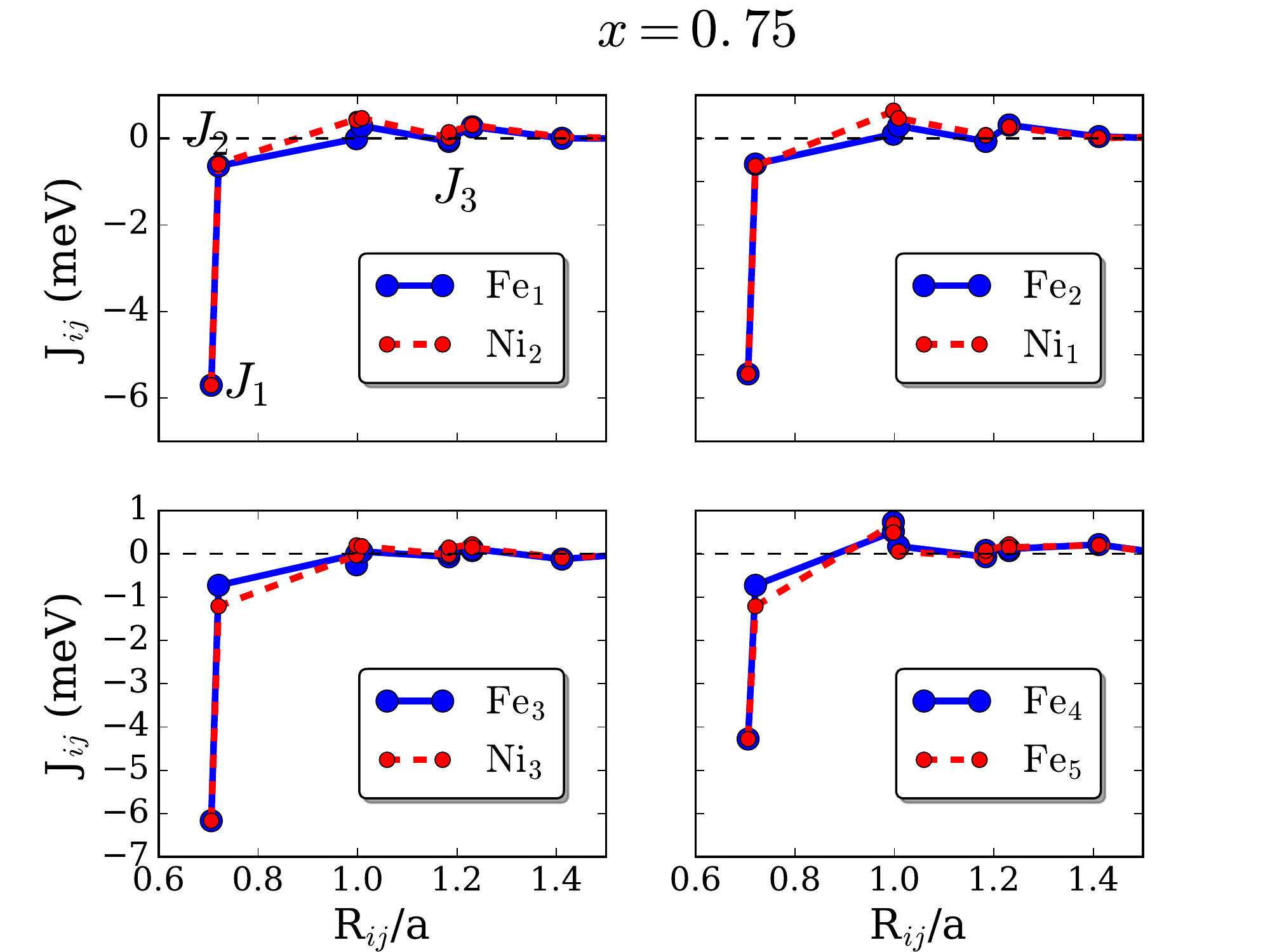} 
 \caption{Exchange parameters between an atom shown in the legend and all its magnetic neighbors as a function of distance for \textit{x} = 0.50 (left panel) and \textit{x} = 0.75 (right panel). For comparison, the atoms belonging to the same plane are shown together.}
\label{fig:J}
\end{figure*}

In the following we compare our experimental results with our theoretical results. As mentioned in the previous section,  the calculations were performed for two Y concentrations, \textit{x} = 0.5 and 0.75, taking into account that the O1-site is vacant, while the O2- and O3-sites are fully occupied ($\delta=1$). A supercell consisting of four formula units of the original unitcell (Sr$_{3}$Fe$_2$O$_7$) is then constructed, e.g., Sr$_{12}$Fe$_8$O$_{28}$. Three of the Fe atoms are replaced by Ni atoms to keep the ratio of Ni to Fe equal to that of the experimental samples. This substitution is done so that the Ni atoms are kept as far from each other as possible, in order to get a homogeneous structure. Figure~\ref{fig:Y075} shows the positions and notations used for the different 3$d$ atoms. The same rule has also been applied when Sr atoms are replaced with Y atoms.

\begin{table}[h]
\centering
{
\caption{\label{tab:table1} Total energy difference ($\Delta$E) with respect to the lowest energy configuration per formula unit (f.u.) as well as the averaged on-site magnetic moment of different spin configurations in the LDA+$U$ approach for $x=0.5$ system. The $+/-$ signs in the spin configuration states refer to the orientations of each 3\textit{d} spin moment, i.e., spin up/down. }
\begin{ruledtabular}
\begin{tabular}{ cccc }
\quad  Spin Configuration \quad   &\multicolumn{1}{c}{$\Delta$E }&\multicolumn{1}{c}{ $\mu_{\textrm{avg}}$ } \\ [1ex]
 \scriptsize{ | Fe$_1$ Fe$_2$ Fe$_3$ Fe$_4$ Fe$_5$ Ni$_1$ Ni$_2$ Ni$_3 \rangle $}   &	\: (eV)/f.u. \: & \:  ($\mu_B$) \:  \\
 \hline 
| $ +\:\: +\:\: +\:\: +\:\: -\:\: -\:\: -\:\: -\:\: \rangle $   &	 0.016 &   2.92   \\
| $ +\:\: -\:\: +\:\: +\:\: -\:\: +\:\: -\:\: -\:\: \rangle $   &	 0.026 &   2.92    \\
| $ +\:\: +\:\: -\:\: +\:\: -\:\: -\:\: -\:\: +\:\: \rangle $   &	 0.000 &   2.92    \\
| $ +\:\: -\:\: -\:\: +\:\: -\:\: +\:\: -\:\: +\:\: \rangle $   &	 0.010 &   2.92   \\
| $ +\:\: -\:\: +\:\: +\:\: -\:\: +\:\: -\:\: +\:\: \rangle $  &	 0.085 &   2.94    \\
| $ +\:\: +\:\: +\:\: -\:\: +\:\: +\:\: +\:\: -\:\: \rangle $ &	 0.113 &	 2.98    \\
| $ +\:\: -\:\: +\:\: +\:\: -\:\: +\:\: +\:\: +\:\: \rangle$  &   	0.193  &   3.00   \\    
| $ +\:\: +\:\: +\:\: +\:\: -\:\: -\:\: -\:\: +\:\: \rangle $  &       0.076  &   2.94    \\
| $ +\:\: +\:\: -\:\: +\:\: +\:\: -\:\: -\:\: +\:\: \rangle $  &	0.274  &   2.93  \\    
| $ +\:\: +\:\: -\:\: +\:\: -\:\: -\:\: -\:\: -\:\: \rangle $    & 0.260  &   2.87    \\[1ex]
\end{tabular}
    \end{ruledtabular}
}

\centering
{
\caption{\label{tab:table2} Site-projected as well as the averaged (avg) on-site magnetic moments of the obtained ground states for two different Y concentrations ($x$).}
    \begin{ruledtabular}
    \begin{tabular}{lccccccccc}
            & \multicolumn{8}{c}{magnetic moment $(\mu_B)$}\\
            \cline{2-10}
        $x$ & Fe$_1$ & Fe$_2$ & Fe$_3$ & Fe$_4$ & Fe$_5$ & Ni$_1$ & Ni$_2$ & Ni$_3$ & avg \\ [1ex]
        \hline
         0.50 &  3.58 & 3.68 & -3.69 & 3.75 & -3.75 & -1.60 & -1.61 & 1.58 & 2.92  \\  [1ex]
         0.75 & 3.75 & 3.78 & -3.68 & 3.71 & -3.71 & -1.54 & -1.52 & 1.52 & 2.94   \\ [1ex] 
    \end{tabular}
    \end{ruledtabular}
}
\end{table}

Consequently, the ground state magnetic configuration as well as the corresponding total energies have been extracted from the self-consistent calculations. For this, we studied the energies of all possible magnetic (spin) configurations, which could be accommodated within the supercell, for each Y concentration. In Table~\ref{tab:table1}, we show some of the magnetic configurations that we considered in our simulations as well as their total energy difference with respect to the ground state ($\Delta$E) and their averaged on-site magnetic moments ($\mu_{\textrm{avg}}$). The magnetic moment of the system mainly arises from 3\textit{d} atoms, i.e., Fe and Ni. The small induced moments associated to other types of atoms (Sr, Y and O) have no significant contribution to the total moment of the system.  Therefore, we define the averaged on-site magnetic moment as the sum over all the moments rising from Fe and Ni divided by the total number of 3$d$ atoms, in this case 8. This makes it possible to be able to compare our theoretical results with experimental total moments in Table~\ref{tab:mag-exp}. Based on this observation, we found that for the \textit{x} = 0.50 Y concentration (and in separate calculations for \textit{x} = 0.75) the magnetic configuration of the system is a \textit{G}-type ferrimagnet with an averaged moment of 2.92~$\mu_B$ (2.94~$\mu_B$). Table~\ref{tab:table2} shows the magnetic moment of each 3\textit{d} atom for two different Y concentrations. One can notice that for the same number of O-vacancies, the increase of Y concentration changes the total moment of the system only slightly while this change is more pronounced for the 3\textit{d} atoms close to the substitute site, e.g., Fe$_1$, Fe$_2$, Ni$_1$ and Ni$_2$.

\begin{figure}[t]
\includegraphics[width=\columnwidth]{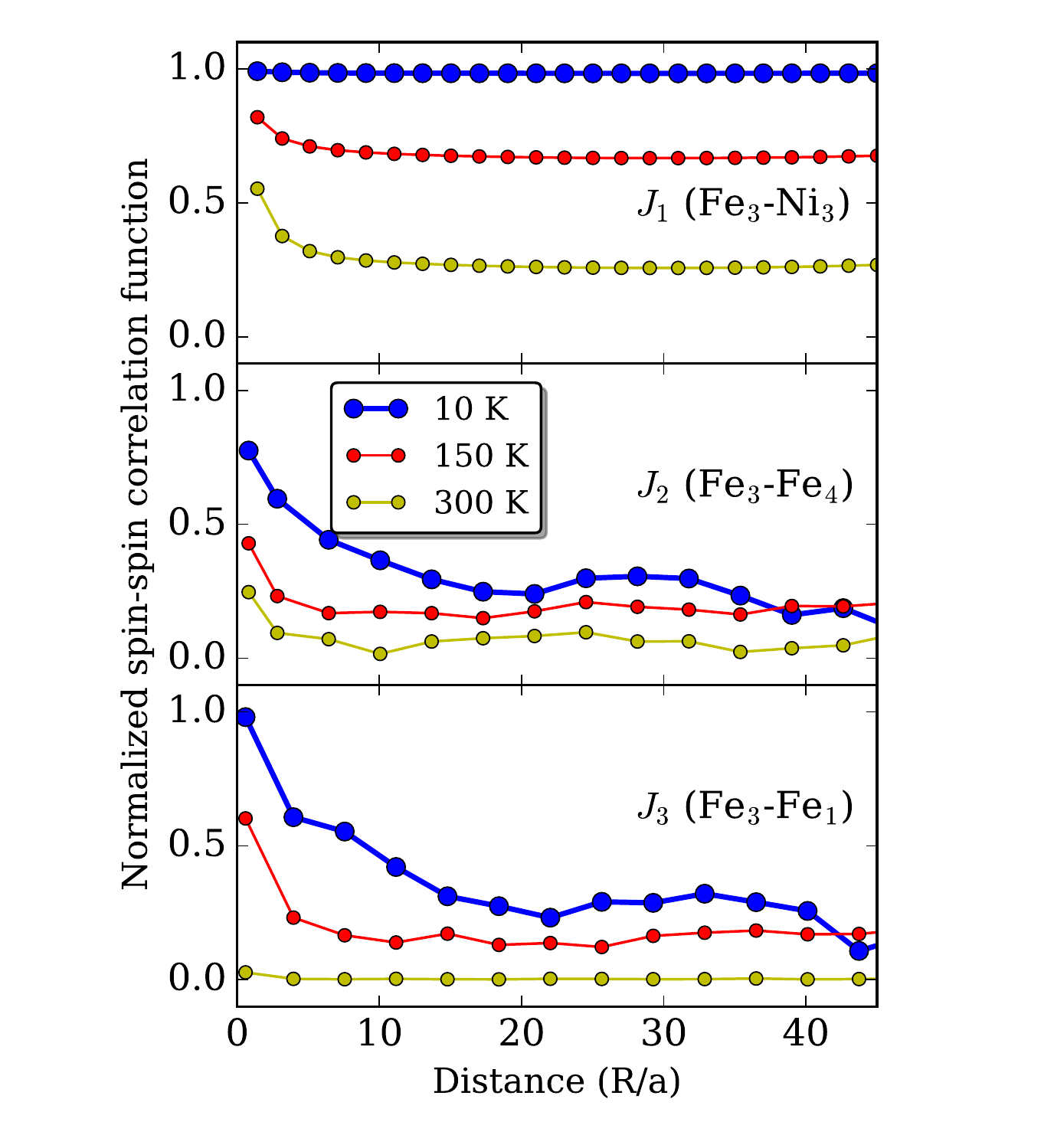}
 \caption{Temperature dependence of the normalized spin-spin correlation function as a function of distance per lattice constant a (R/a) with the intra-layer ($J_1$) and inter-layer ($J_2$ and $J_3$) exchange couplings for the case of interactions between Fe$_3$ and its neighbors. Simulations are performed for the alloy concentration with $x=0.5$. }
\label{fig:correlation}
\end{figure}

We now proceed with an analysis of the interatomic exchange interactions ($J_{ij}$). Figure~\ref{fig:J} shows the exchange parameters between each magnetic atom in the system with all other neighbors as a functions of distance, for two different Y concentrations: left panel for \textit{x} = 0.50 and right panel for \textit{x} = 0.75. An important point in these plots is that the nearest intra-layer ($J_1$) and inter-layer ($J_2$) exchange interactions are AFM which is consistent with the measured $G$-type AFM order in these compounds. Additionally, the 3D ordering has been observed in the systems due to the non-zero inter-layer exchange interactions ($J_2$ and $J_3$) while the dipole-dipole interaction apeared to be quite weak. Similar situation has been reported before for SrFeO$_2$~\cite{layered-3,Hayashi} and Sr$_3$F$_2$O$_5$~\cite{Hayashi}, with the 3D ordering as a result of the finite inter-layer exchange interaction. Contrary, in many other layered-perovskites the 3D ordering could only be stabilized by the help of magnetic dipole-dipole interaction~\cite{Koo,Sr2TcO4,Ru2MnF4-1,Ru2MnF4-2,Ru2MnF4-3}. Among them is the two different studies on Sr$_3$F$_2$O$_5$ which reported the 3D ordering due to the inter-layer exchange~\cite{Hayashi} or dipole-dipole interaction~\cite{Koo}. 
We also note that from the self-consistent spin-wave theory for quasi-2D systems, it is shown that the presence of small inter-layer exchange couplings are important and may provide 3D ordering~\cite{berez-1,berez-2,3d-j}. This 3D ordering can exist in low temperatures but will undergo a transition to 2D ordering with increasing the thermal fluctuations~\cite{cross-svis}.

The coherence between the atomic moments in different layers is present at low temperatures and slowly disappears close to the $T_N$ due to the thermal fluctuations. This is concluded by the inspection of the spin-spin correlation functions along different directions by considering only the atoms with the most relevant exchange couplings ($J_1$, $J_2$ and $J_3$). The results for the case of Fe$_3$ atom in $x=0.50$ are shown in Fig.~\ref{fig:correlation}. Similar spin-spin correlation function has also been obtained for $x=0.75$ Y (not shown here). As can be seen, at low temperatures where the exchange energy is larger than the thermal energy, both in-plane and out-of-plane correlation functions are finite and the materials exhibits a 3D type of ordering. If the thermal fluctuations become larger, the correlation decays faster, especially between layers that have largest distance between them, i.e. between layers that contain Fe$_3$ and Fe$_1$ atoms (see Fig.~\ref{fig:Y075} for the geometry). At 150 K the correlation between atoms in these layers is weak, and at 300 K it is vanishingly small (Fig.~\ref{fig:correlation}). However, the spin-spin correlation within the plane (corresponding to the larger $J_1$ coupling) is finite even above $T_N$. One way to describe this, is that at low temperature the material has 3D ordering but for elevated temperatures, a transition is made to 2D order.  This is discussed further in the appendix. 

We have evaluated the Ne\'el temperature ($T_N$) based on the obtained exchange parameters. The ordering temperature was determined as the temperature where the specific heat C($T$) exhibits a maximum. Using classical Monte Carlo simulations as implemented in the UppASD code, we have obtained an ordering temperature of 260 K and 310 K for \textit{x} = 0.50 and 0.75, respectively.
These results are in good agreement with our experimentally measured ordering temperatures; 275 K and 310 K for \textit{x} = 0.50 and 0.75, respectively. Excluding the interlayer exchange parameters in the calculations of the ordering temperature decreased this value to 260 K for \textit{x} = 0.75, i.e., a reduction of ordering temperature with 16\%, which illustrates the important role of the interlayer exchange for the proper determination of $T_N$. The effect of the on-site magnetocrystalline anisotropy on $T_N$ was quite marginal (data not shown) which confirms previous studies that exchange interaction is the leading term for the value of ordering temperature~\cite{layered-3}. We note also that the ordering temperature is in a thermal range that is consistent with the energy of the exchange parameter in Fig.~\ref{fig:J}. Finally, we should mention that including dipole-dipole interactions in our calculations did not introduce any noticeable change of the 3D ordering in the system.

\section{\label{sec:level5}Conclusion}
The magnetic properties of Sr$_{3-x}$Y$_{x}$(Fe$_{1.25}$Ni$_{0.75}$)O$_{7-\delta}$ ($0 \leq x \leq 0.75$ and $0< \delta <1$) as obtained from SQUID magnetometry and NPD have been presented. The results are in good agreement with the $\textit{ab initio}$ calculations based on DFT+$U$ that provide magnetic moments and Heisenberg exchange parameters. These parameters specify an effective spin-Hamiltonian that allows for an evaluation of the ordering temperature via Monte Carlo simulations. Both the calculated magnetic moments as well as the ordering temperature are in acceptable agreement with observations. Based on our observations, we conclude that all the studied samples are antiferromagnetically ordered at low temperatures showing an increase in the ordering temperature with Y concentration and O occupancy. Due to no visible asymmetry in the magnetic Bragg profiles, the NPD results strongly indicate 3D magnetic ordering. As opposed to many other Ruddlesden-Popper systems reporting 3D magnetic ordering stabilized by magnetic dipole-dipole interaction, the observed 3D AFM ordering in the samples studied here is attributed to the inter-layer exchange coupling that our theoretical calculations show is not negligible. As a result of the finite interlayer exchange interaction, we observe significant spin-spin correlations among magnetic moments in different crystallographic layers, even at temperatures as high as 150 K. However, at sufficiently high temperature, thermal fluctuations swamp any effect of the $J_3$ exchange interaction, and the material becomes a 2D magnetic system. For the $x=0.5$ system, the simulations indicate that the 3D to 2D crossover occurs at a temperature between 150 K and the magnetic ordering temperature (cf. Appendix). For the real samples, the NPD results indicate that if there is a 3D to 2D crossover, it occurs close to the magnetic ordering temperature. The intralayer spin-spin correlation is found to be finite, even at temperatures larger than the ordering temperature. 

\begin{figure*}[tp]
 \includegraphics[width=0.27\textwidth]{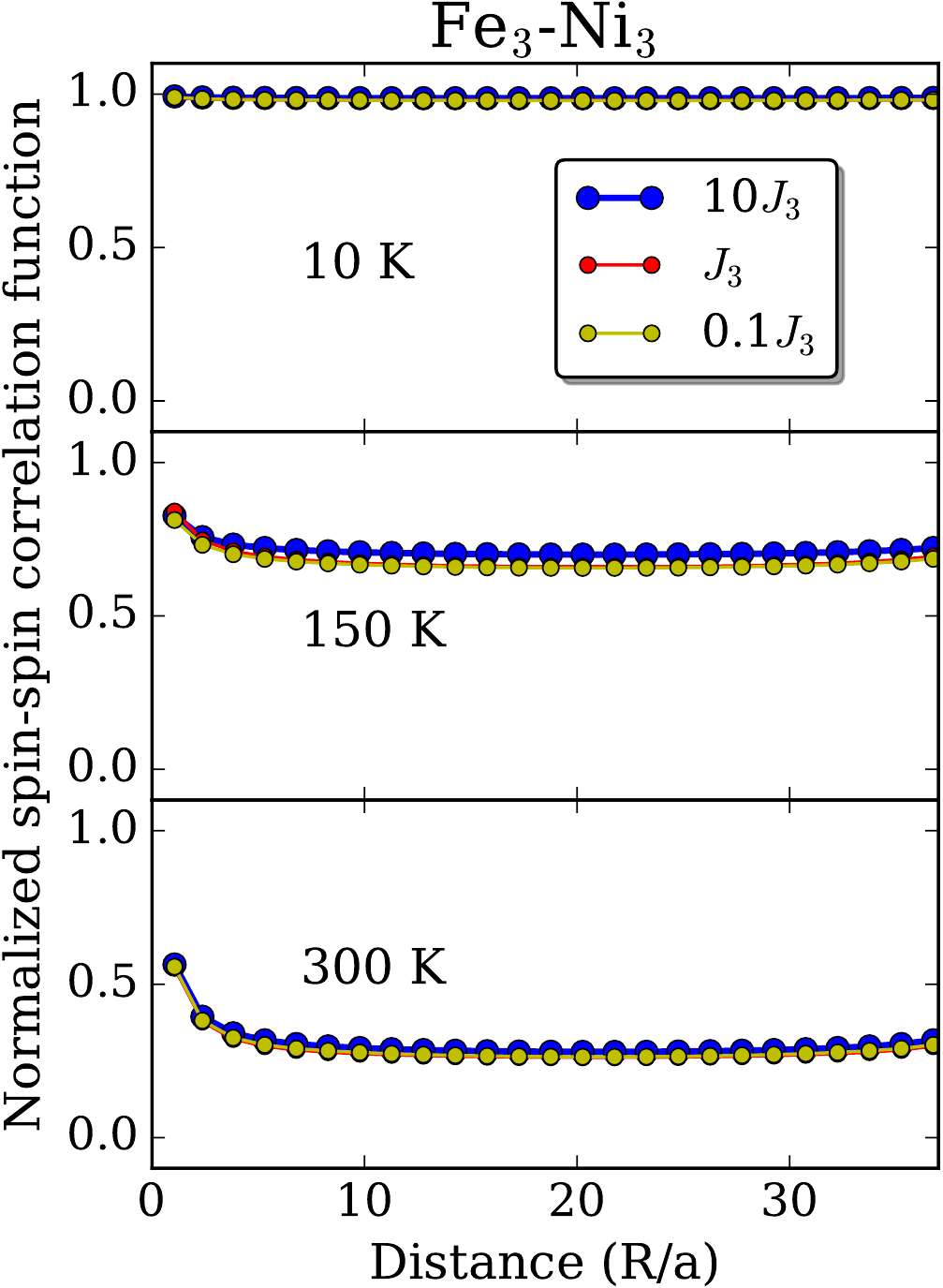} \qquad \includegraphics[width=0.25\textwidth]{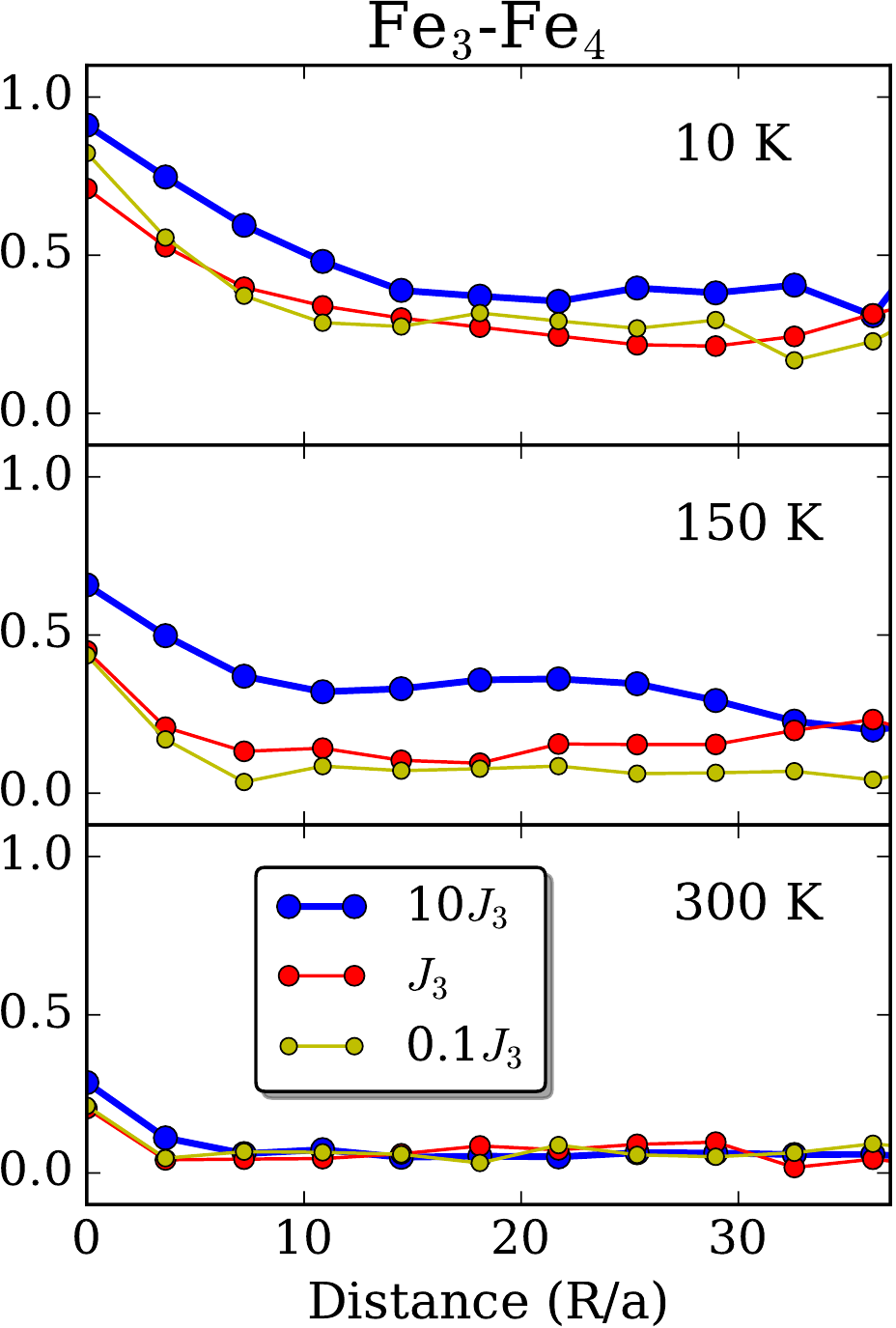} \qquad \includegraphics[width=0.25\textwidth]{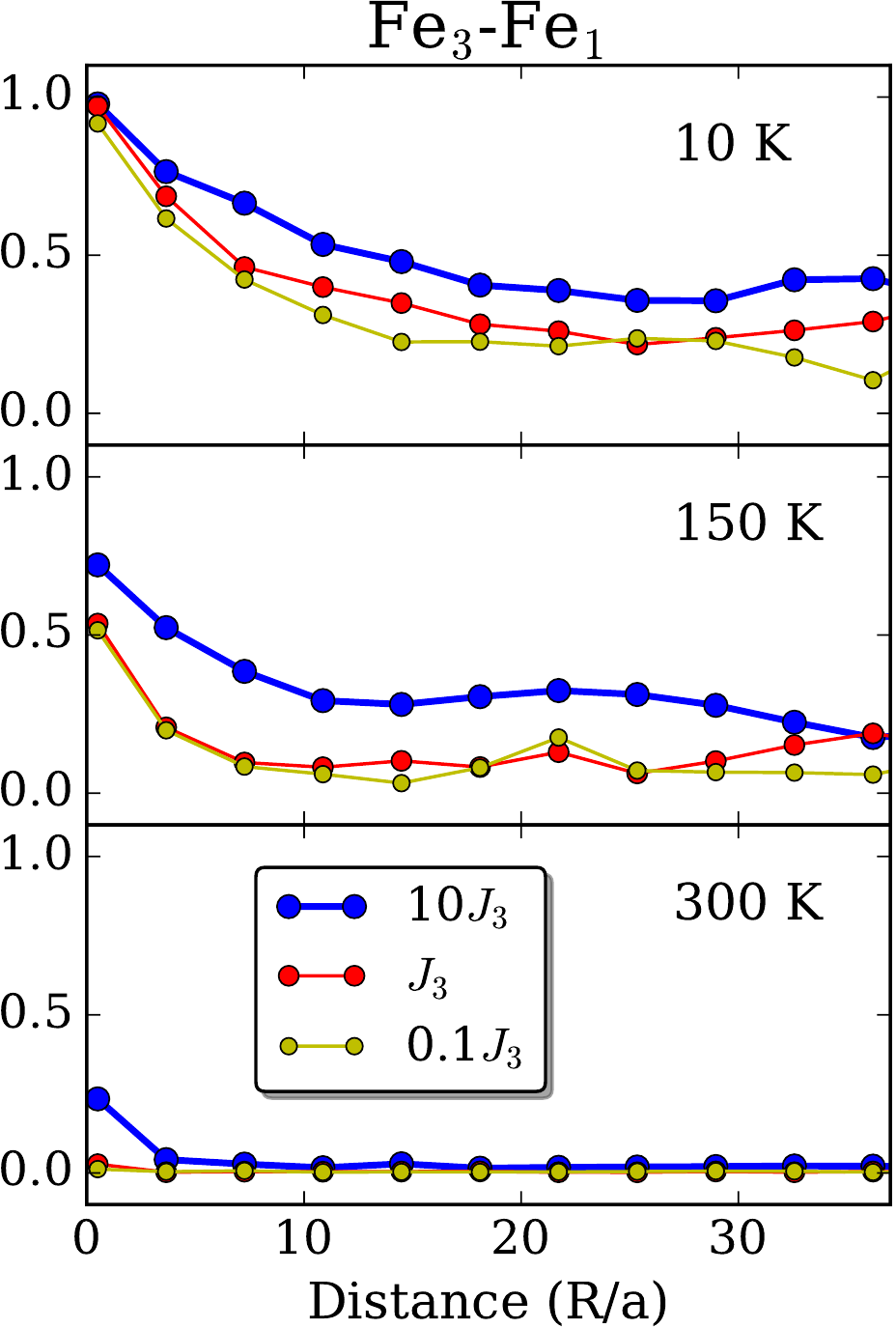}
 \caption{Normalized spin-spin correlation function as a function of distance per lattice constant a (R/a) for the case of interactions between Fe$_3$ and its neighbors. The simulations are performed for the $x=0.5$ Y concentration for various strengths of $J_3$ parameter. }
\label{fig:scaled-j3}
\end{figure*}

\begin{figure}[t]
 \includegraphics[width=\columnwidth]{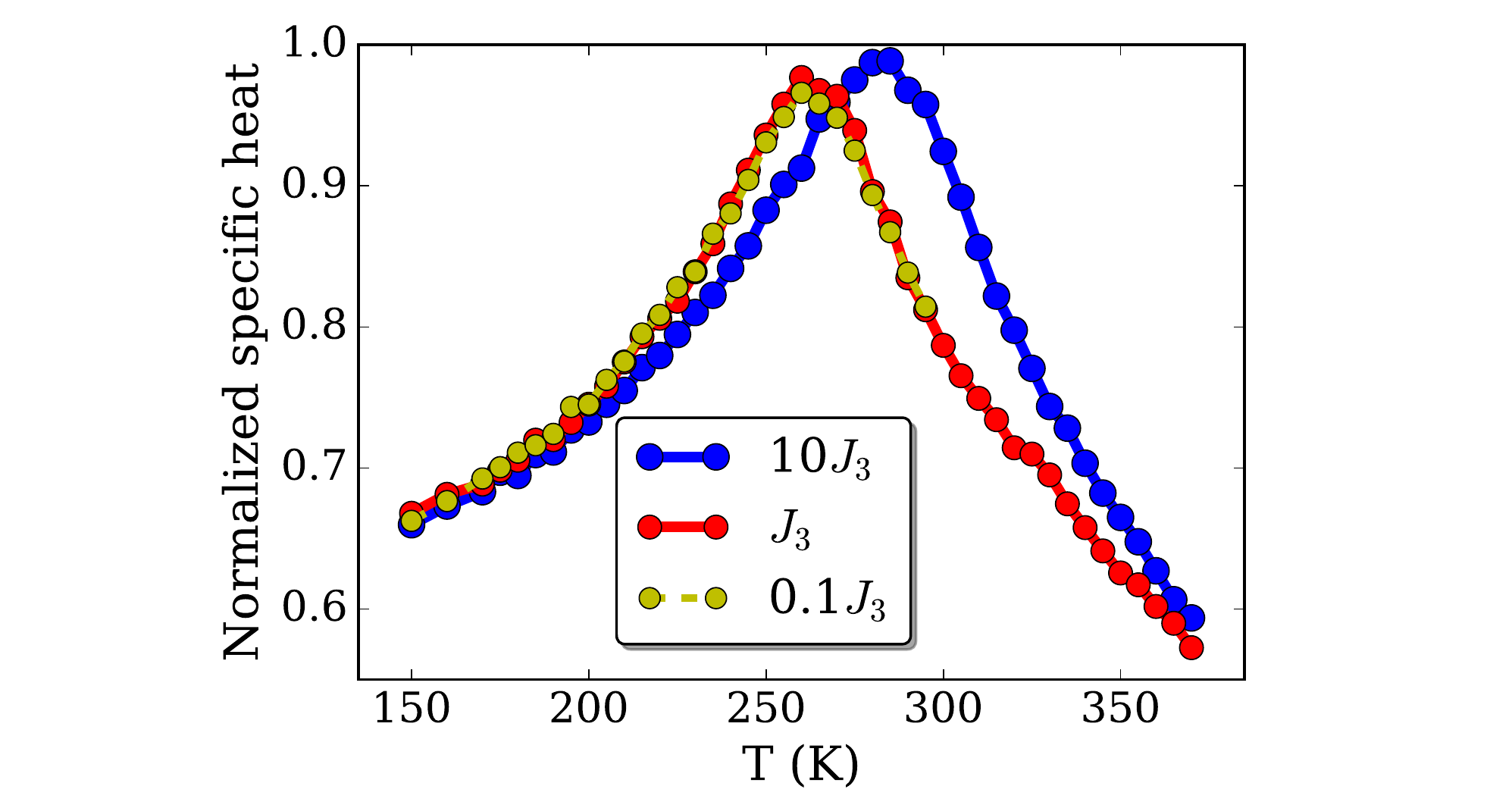}
 \caption{Normalized specific heat as a function of temperature for various strengths of $J_3$ parameter, for the $x=0.5$ Y concentration.}
\label{fig:specific}
\end{figure}

\section{\label{sec:level6}Acknowledgment}

The computer simulations are performed on computational resources provided by NSC, UPPMAX and PDC centre for high performance computing (PDC-HPC) allocated by the Swedish National Infrastructure for Computing (SNIC). O. E. acknowledges support from the Swedish Research Council (VR), eSSENCE and the KAW foundation. S. K. acknowledges Patrik Thunstr\"om for fruitful discussion and Johan Sch\"ott for his help in matplotlib. We acknowledge Dr. Naveen Kumar Chogondahalli Muniraju from Spallation Neutron Source for his help with the NPD measurements.

\appendix

In order to investigate further the influence of the $J_3$ parameter on the  nature of magnetic order at different temperatures, we performed simulations of the spin-spin correlation function between atoms in different layers for various strengths of the $J_3$ parameter. The simulations were done using the value of 
 $J_3$ as obtained from the first principles theory, as well as a ten fold increased and a ten fold decreased value. As shown in Fig.~\ref{fig:scaled-j3}, the correlations between atoms in the plane containing Fe$_3$ and Ni$_3$ atoms, is basically insensitive to the strength of $J_3$, at least in the range of values considered here. This is expected since it is the $J_1$ parameter that primarily determines these correlations. The situation is different for the other correlations illustrated in Fig.~\ref{fig:scaled-j3}. At low temperatures, a modification of $J_3$ is seen to influence the spin-spin correlations in an expected way; an increased value increases also the spin-spin correlations. However, at sufficiently high temperature, e.g. 300 K, thermal fluctuations swamp any effect of the  $J_3$ exchange interaction, and the material becomes a 2D magnetic system. We have also evaluated the specific heat for scaled values of $J_3$. In Fig.~\ref{fig:specific}, the peak in the specific heat gives an estimate of the ordering temperature, and it can be seen that a tenfold increase of $J_3$ influences the ordering temperature somewhat, while a tenfold decrease of $J_3$ results in an ordering temperature that is the same as the unscaled value. We conclude from this exercise that the unscaled value of $J_3$ corresponds to a 2D magnet at the magnetic ordering temperature, while an increase of $J_3$ with 10 provides a weak 3D system. Moreover, the spin-spin correlations shown in Fig.~\ref{fig:scaled-j3} indicate that there is a 3D to 2D crossover at some  temperature between 150 K and the magnetic ordering temperature. The experimental NPD results show that the  magnetic state for the $x=0.75$ sample is still 3D at 300 K, indicating that if there is a 3D to 2D crossover also for the real samples, it occurs close to the magnetic ordering temperature.

\bibliographystyle{apsrev4-1}

\end{document}